\documentclass[runningheads]{llncs}
\usepackage{graphicx}
\usepackage{hyperref}
\usepackage{float}
\usepackage{siunitx}
\usepackage{booktabs}

\usepackage[dvipsnames]{xcolor}

\newcolumntype{P}[1]{>{\centering\arraybackslash}p{#1}}

\hypersetup{
    colorlinks = true,
    linkcolor = blue,
    anchorcolor = blue,
    citecolor = blue,
    filecolor = blue,
    urlcolor = blue
}

\begin{document}
\title{Optimized U-Net for Brain Tumor Segmentation}
\author{Michał Futrega, Alexandre Milesi, Michał Marcinkiewicz, Pablo Ribalta}
\institute{NVIDIA, Santa Clara CA 95051, USA\\
\email{\{mfutrega,alexandrem,michalm,pribalta\}@nvidia.com}}

\maketitle           
\begin{abstract}
We propose an optimized U-Net architecture for a brain \mbox{tumor} segmentation task in the BraTS21 challenge. To find the \mbox{optimal} model architecture and the learning schedule, we have run an \mbox{extensive} ablation study to test: deep supervision loss, Focal loss, decoder \mbox{attention}, drop block, and residual connections. Additionally, we have searched for the optimal depth of the U-Net encoder, number of convolutional \mbox{channels} and post-processing strategy. Our method won the validation phase and took third place in the test phase. We have open-sourced the code to reproduce our BraTS21 submission at the NVIDIA Deep \mbox{Learning} Examples GitHub Repository\footnote{\url{https://github.com/NVIDIA/DeepLearningExamples/blob/master/PyTorch/Segmentation/nnUNet/notebooks/BraTS21.ipynb}}.
\end{abstract}

\section{Introduction}

One of the most challenging problems in medical image processing is automatic brain tumor segmentation. Obtaining a computational model capable of surpassing a trained human-level performance would provide valuable assistance to clinicians and would enable a more precise, reliable, and standardized approach to disease detection, treatment planning and monitoring. Gliomas are the most common type of brain tumors in humans~\cite{introduction_glioma}. Their accurate segmentation is a challenging medical image analysis task due to their variable shape and appearance in multi-modal magnetic resonance imaging (MRI). Manual segmentation of such brain tumors requires a great deal of medical expertise, is time-consuming, and prone to human error. Moreover, the manual process lacks consistency and reproducibility, which negatively affects the results and can ultimately lead to incorrect prognosis and treatment.

The rapid progress in development of deep learning (DL) algorithms shows great potential for application of deep neural networks (DNNs) in computer-aided automatic or semi-automatic methods for medical data analysis. The drastic improvements of convolutional neural networks (CNNs) resulted in models being able to approach or surpass the human level performance in plethora of applications, such as image classification~\cite{introduction_imagenet} or microscope image segmentation~\cite{introduction_deepem3d}, among many others. DL-based models are great candidates for brain tumor segmentation, as long as sufficient amount of training data is supplied. The Brain Tumor Segmentation Challenge (BraTS) provides a large, high-quality dataset consisting of multi-modal MRI brain scans with corresponding segmentation masks~\cite{data1,data2,data3,data4,data5}.

The state-of-the-art models in brain tumor segmentation are based on the encoder-decoder architectures, with U-Net~\cite{introduction_unet} being the most popular for medical image segmentation, based on the citations number. In recent years, U-Net-like architectures were among top submissions to the BraTS challenge. For instance, in 2018, Myronenko \textit{et al.,} modified a U-Net model by adding a variational autoencoder branch for regularization~\cite{introduction_myronenko}. In 2019, Jiang \textit{et al.,} employed a two-stage U-Net pipeline to segment the substructures of brain tumors from coarse to fine~\cite{introduction_jiang}. In 2020, Isensee \textit{et al.,} applied the nnU-Net framework~\cite{introduction_nnunet} with specific BraTS designed modifications regarding data post-processing, region-based training, data augmentation, and minor modifications to the nnU-Net pipeline~\cite{introduction_nnunet_brats}.

Those achievements prove that well-designed U-Net based architectures have the ability to perform very well on tasks such as brain tumor segmentation. In order to design a competitive solution for challenges like BraTS21, both optimal neural network architecture and training schedule has to be selected. However, there exists a plethora of U-Net variants, for example: Attention U-Net~\cite{atten}, Residual U-Net~\cite{resnet}, Dense U-Net~\cite{dense}, Inception U-Net~\cite{inception}, U-Net++~\cite{unet++}, SegResNetVAE~\cite{introduction_myronenko} or UNETR~\cite{introduction_unetr}, just to name a few. A wide range of U-Net architectures makes the selection of the optimal one a difficult task. Furthermore, once the neural network architecture is selected, designing a proper training schedule is critical for getting optimal performance. Designing a training schedule is associated with selecting optimal components, such as a loss function, data augmentation strategy, learning rate and its schedule, number of epochs to train, and many more. Also, it is not trivial to decide which model extensions to add, for example, deep-supervision~\cite{ds} or drop-block~\cite{drop}.

The fact that datasets for medical image segmentation are small (usually around 100 examples), and there is no common benchmark for measuring improvements of different architecture tweaks, often makes such comparisons unreliable. However, the dataset released for BraTS21 provides 2,040 examples (respectively 1251, 219, 570 examples in the training, validation, and test set), which makes it the largest dataset for medical image segmentation at the moment, and a perfect candidate to measure performance improvements for different U-Net variants.

In this paper, we have run extensive ablation studies to select both an optimal U-Net variant and training schedule for the BraTS21 challenge. We have tested U-Net\cite{introduction_unet}, Attention U-Net~\cite{atten}, Residual U-Net~\cite{resnet}, SegResNetVAE~\cite{introduction_myronenko} and UNETR~\cite{introduction_unetr} for \mbox{U-Net} variants, and experimented with Deep Supervision~\cite{ds}, Drop-Block~\cite{drop}, and different loss functions (combinations of Cross Entropy, Focal, and Dice). Furthermore, we have optimized our model further by increasing the encoder depth, adding one-hot-encoding channel for the foreground voxels to the input data, and increasing the number of convolutional filters.

\newpage
\section{Method}\label{ch:method}
\subsection{Data}\label{sub:data}

\begin{figure}[t!]
\centering
\includegraphics[scale=0.74]{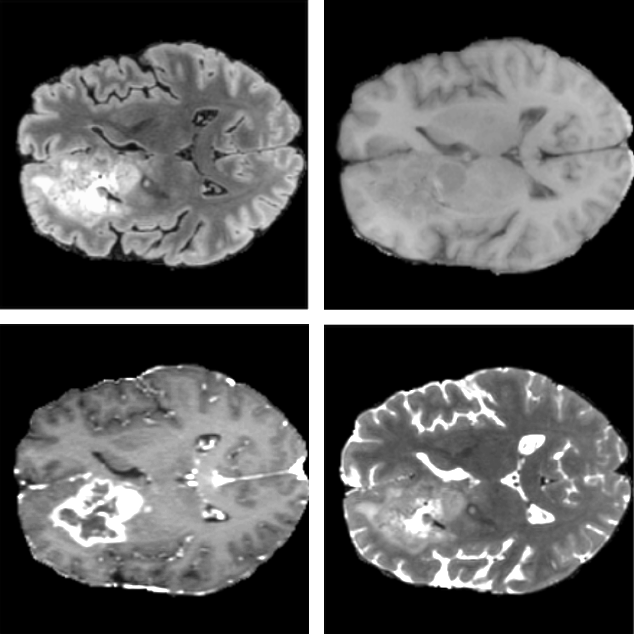}
\caption{Example with ID 00000 from the BraTS21 training dataset. Each subplot presents a different modality. From top left to bottom right: FLAIR, T1, T1Gd T2.}
\label{fig:dataset}
\end{figure}

The training dataset provided for the BraTS21 challenge~\cite{data1,data2,data3,data4,data5} consists of $1,251$ brain MRI scans along with segmentation annotations of tumorous regions. The 3D volumes were skull-stripped and resampled to $1$ mm\textsuperscript{3} isotropic resolution, with dimensions of $(240, 240, 155)$ voxels. For each example, four modalities were given: native (T1), post-contrast T1-weighted (T1Gd), T2-weighted (T2), and T2 Fluid Attenuated Inversion Recovery (T2-FLAIR). Example images of each modality are presented on Fig. \ref{fig:dataset}. Segmentation labels were annotated manually by one to four experts. Annotations consist of four classes: enhancing tumor (ET), peritumoral edematous tissue (ED), necrotic tumor core (NCR), and background (voxels that are not part of the tumor).

\subsection{Data Preprocessing and Augmentations}\label{sub:prep}

Each example of the BraTS21 dataset consists of four NIfTI~\cite{nifti} files with different MRI modalities. As a first step of data pre-processing, all four modalities were stacked such that each example has a shape of $(4, 240, 240, 155)$ (input tensor is in the (C, H, W, D) layout, where C-channels, H-height, W-width and D-depth). Then redundant background voxels (with voxel value zero) on the borders of each volume were cropped, as they do not provide any useful information and can be ignored by the neural network. Subsequently, for each example, the mean and the standard deviation were computed within the non-zero region for each channel separately. All volumes were normalized by first subtracting the mean and then divided by the standard deviation. The background voxels were not normalized so that their value remained at zero. To distinguish between background voxels and normalized voxels which have values close to zero, an additional input channel was created with one-hot encoding for foreground voxels and stacked with the \mbox{input data.}

Data augmentation is a technique that alleviates the overfitting problem by artificially extending a dataset during the training phase. To make our method more robust, the following data augmentations were used during training phase:

\begin{enumerate}
 \item \textbf{Biased crop}: From the input volume, a patch of dimensions $(5, 128, 128, 128)$ was randomly cropped. Additionally, with probability of $0.4$ the patch selected via random biased crop is guaranteed that some foreground voxels (with positive class in the ground truth) are present in the \mbox{cropped region.} \vspace{1mm}
 \item \textbf{Zoom}: With probability of $0.15$, a random value is sampled uniformly from $(1.0, 1.4)$ and image size is resized to its original size times the sampled value with the cubic interpolation, while the ground truth with the nearest neighbour interpolation. \vspace{1mm}
 \item \textbf{Flips}: With probability of $0.5$, for each $x, y, z$ axis independently, volume was flipped along that axis. \vspace{1mm}
 \item \textbf{Gaussian Noise}: With probability of $0.15$, random Gaussian noise with mean zero and standard deviation sampled uniformly from $(0, 0.33)$ is sampled for each voxel and added to the input volume. \vspace{1mm}
 \item \textbf{Gaussian Blur}: With probability of $0.15$, Gaussian blurring with standard deviation of the Gaussian Kernel sampled uniformly from $(0.5, 1.5)$ is applied to the input volume. \vspace{1mm}
 \item \textbf{Brightness}: With probability of $0.15$, a random value is sampled uniformly from $(0.7, 1.3)$ and then input volume voxels are \mbox{multiplied by it.} \vspace{1mm}
 \item \textbf{Contrast}: With probability of $0.15$, a random value is sampled uniformly from $(0.65, 1.5)$ and then input volume voxels are multiplied by it and clipped to their original value range.
\end{enumerate}

\subsection{Model architecture}\label{sub:model}

In order to select the most optimal neural network architecture, we have run ablation studies for the following models: U-Net~\cite{introduction_unet}, Attention U-Net~\cite{atten}, Residual U-Net~\cite{resnet}, SegResNetVAE~\cite{introduction_myronenko} and UNETR~\cite{introduction_unetr}. Below, we present a short description of each model.

\newpage
\subsubsection{U-Net} 
\begin{figure}[H]
\centering
\includegraphics[scale=0.22]{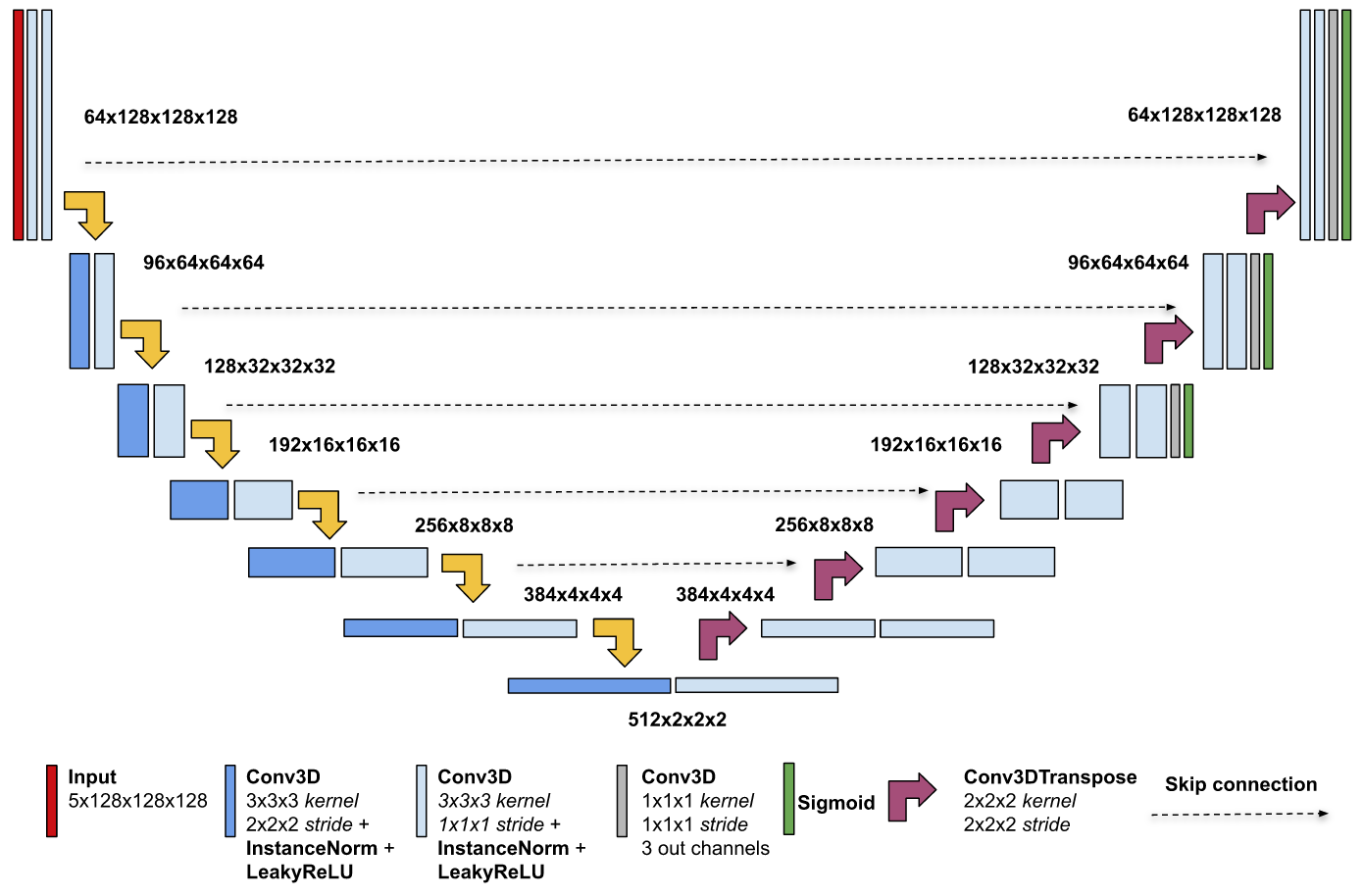}
\caption{U-Net architecture. The encoder is transforming the input by reducing its spatial dimensions, and then the decoder is upsampling it back to the original input shape. Additional two output heads are used for deep supervision loss (green bars).}
\label{fig:unet}
\end{figure}

\cite{introduction_unet} architecture (shown in the Fig. \ref{fig:unet}) is characterised by a symmetric U-shape, and can be divided into two parts, i.e., encoder and decoder. The first part is the contracting path (encoder) which is transforming the input volume into lower dimensional space. The encoder has a modular structure consisting of repeating convolution blocks. Each block has two smaller blocks of transformations (dark and light blue blocks on the Fig. \ref{fig:unet}). The first smaller block is reducing the spatial dimensions of the input feature map by a factor of two via convolutional layer with kernels 3x3x3 and stride 2x2x2, then instance normalization and Leaky ReLU activation with negative slope if 0.01 are applied (dark blue block). Next feature map is transformed with almost the same set of operations except that the convolutional layer has stride 1x1x1 (light blue).

After the spatial dimensions of the feature map are transformed to the size of 2x2x2, then the decoder part starts. The decoder also has a modular structure, but its goal is to increase the spatial dimensions by reducing the encoder feature map. The block in the decoder is built from three smaller blocks. The first one is transposed convolution with kernels 2x2x2 and stride 2x2x2, which is increasing the spatial dimensions of the feature map by a factor of two. Then upsampled feature map is concatenated with encoder feature map from the equivalent spatial level and then transformed by two identical blocks with convolutional layer with kernels 3x3x3 and stride 1x1x1, instance normalization and Leaky ReLU activation with negative slope if 0.01 are applied (light blue). Additionally, deep-supervision (Subsection \ref{sub:ds}) can be used, which is computing loss functions for outputs from lower decoder levels.

\subsubsection{SegResNetVAE} 
\begin{figure}[H]
\centering
\includegraphics[scale=0.3]{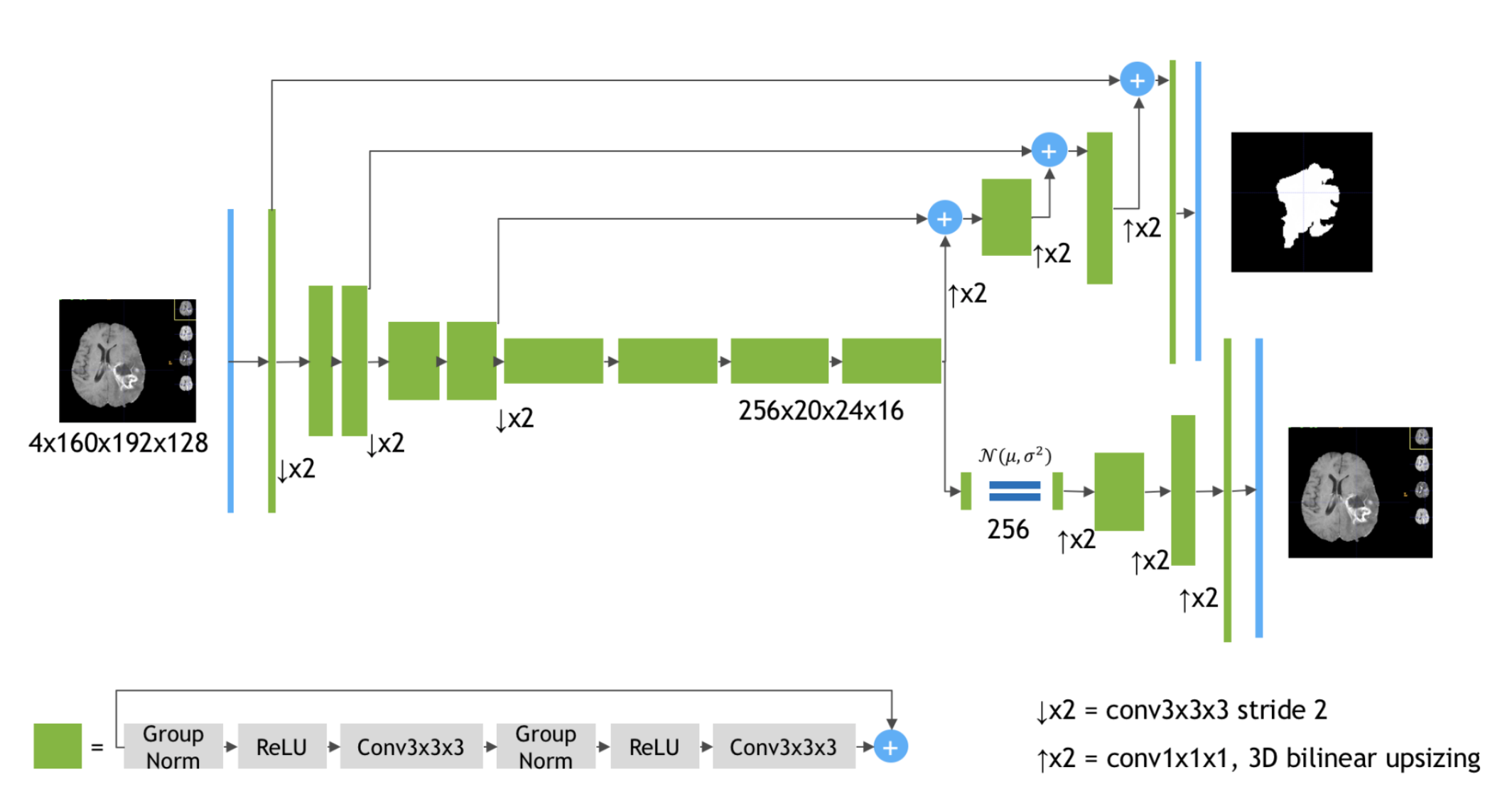}
\caption{SegResNetVAE architecture. Each green block is a ResNet-like block with the group normalization. The VAE branch reconstructs the input image into itself, and is used only during training to regularize the shared encoder. Image from \cite{introduction_myronenko}.}
\label{fig:segres}
\end{figure}

\cite{introduction_myronenko} is a residual U-Net with autoencoder (shown in the \mbox{Fig. \ref{fig:segres})} regularization that has won the BraTS 2018 challenge and modifies U-Net by designing new architecture for encoder blocks and by adding a variational autoencoder (VAE)~\cite{vae} branch in the decoder, which reconstructs the input and has a regularization effect.

The encoder part uses ResNet like blocks, where each block consists of two
convolutions with group normalization and ReLU activation, followed by additive identity skip connection. The decoder structure is similar to the encoder, but only with a single block per each spatial level. Each decoder block begins with reducing the number of channels by a factor of 2 (with 1x1x1 convolution) and doubling the spatial dimension (using 3D bilinear), followed by an addition with encoder feature map from the equivalent spatial level.

In the VAE branch in the decoder, first the feature map from the bottleneck is reduced into a low dimensional space of 256 (128 to represent mean, and 128 to represent std). Then, a sample is drawn from the Gaussian distribution with the given mean and std, and reconstructed into the input image dimensions following the same architecture as the decoder. \vspace{-5mm}

\vspace{-10mm}
\subsubsection{UNETR} 
\begin{figure}[H]
\centering
\includegraphics[scale=0.28]{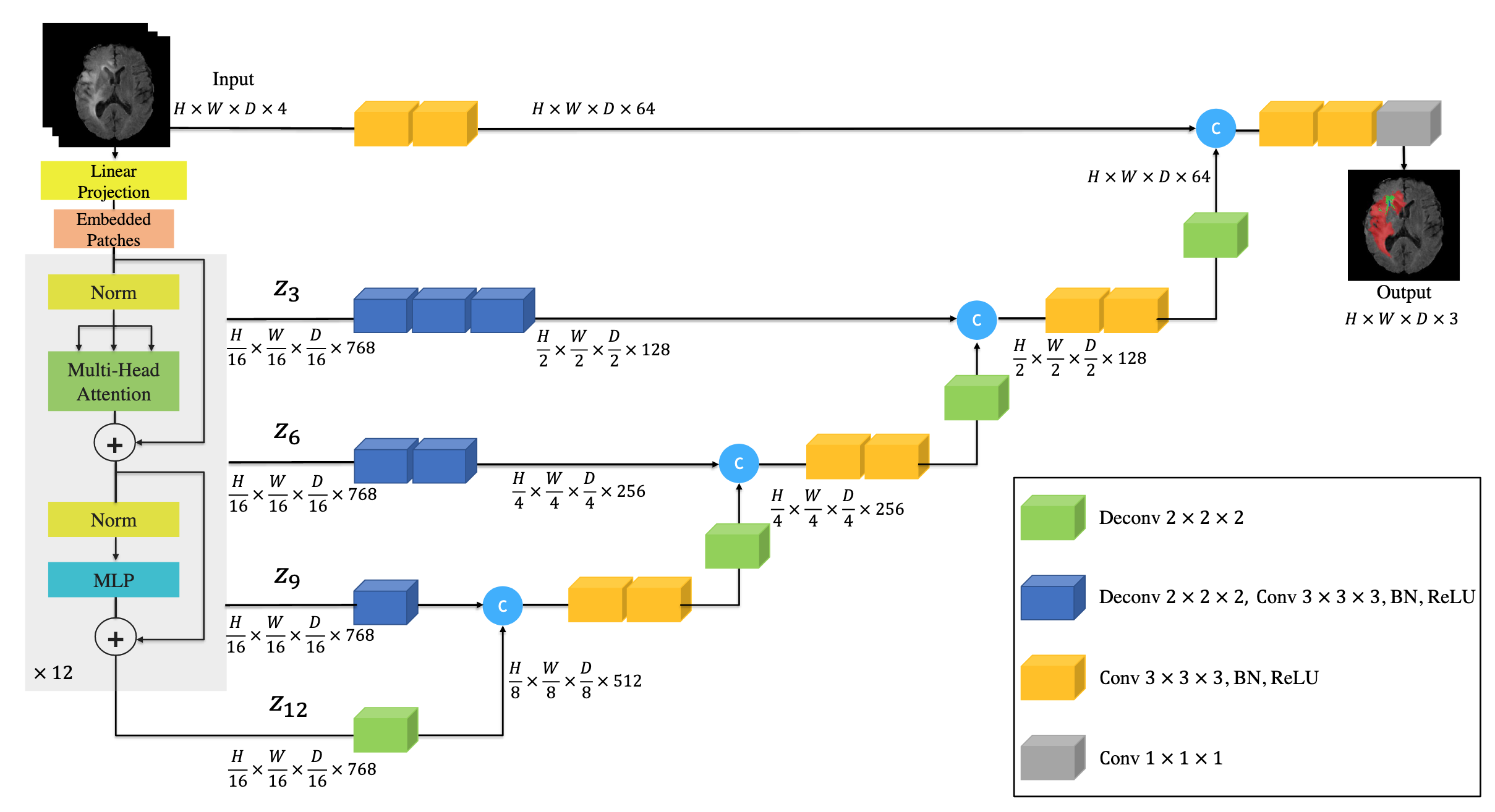}
\caption{UNETR architecture. Instead of using 3D convolution in the encoder, UNETR is transforming the input volume via multi-head self-attention blocks known from the Transformer model. Image from \cite{introduction_unetr}.}
\label{fig:unetr}
\end{figure}

\cite{introduction_unetr} architecture (shown in the Fig. \ref{fig:unetr}) is a generalization of Vision Transformer (ViT)~\cite{vit} to the 3D convolutions---it replaces the 3D convolutions in the encoder with multi-head self-attention~\cite{transformer}. To convert a 3D input volume into an input for a multi-head self-attention it is divided into a sequence of uniform non-overlapping patches (with 16x16x16 shape) and projected into an embedding space (with 768 dimensions) using a linear layer, and added with a positional embedding. Such input is then transformed by a multi-head self-attention encoder.

\subsubsection{Attention U-Net} 
\begin{figure}[H]
\centering
\includegraphics[scale=0.51]{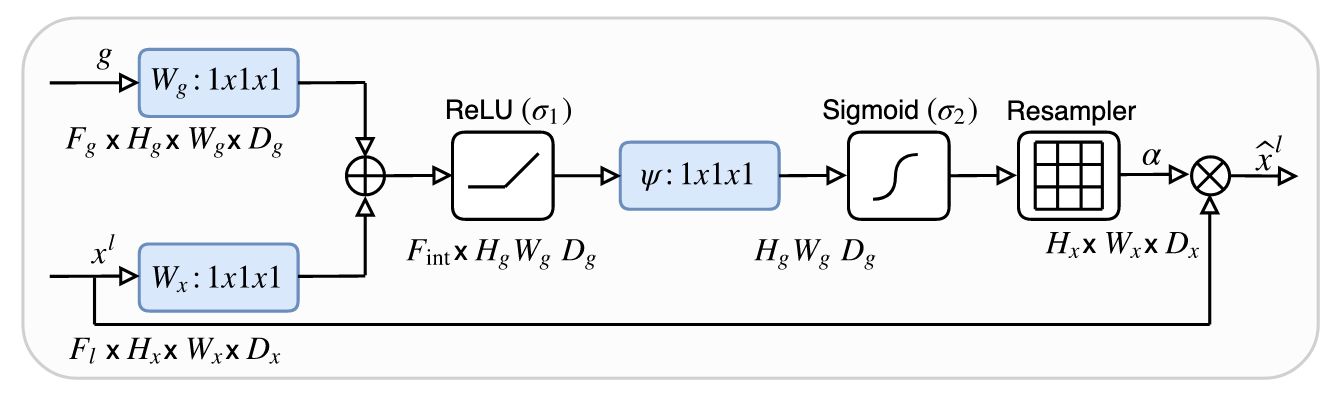}
\caption{The architecture of the attention gate. Input features ($x^l$) are multiplied by attention weights ($\alpha$). To compute $\alpha$, input features ($x^l$), and feature map from corresponding encoder level are first transformed by 1x1x1 convolution, and the summed. Next, ReLU activation and another 1x1x1 convolution are applied. Finally, attention weights are upsampled with trilinear interpolation. Image from \cite{atten}.}
\label{fig:attenunet}
\end{figure}

\cite{atten} is extending base U-Net by adding an attention gate (shown in the Fig. \ref{fig:attenunet}) in the decoder part. Attention gate is transforming the feature map from the encoder before the concatenation in the decoder block. It learns which regions of the encoder feature map are the most important, considering the context of the feature map from the previous decoder block. This is achieved by multiplication of the encoder feature map with the weights computed by the attention gate. The weight values are in the (0, 1) range and represent the attention level that the neural network is paying to a given pixel.

\subsubsection{Residual U-Net} 
\begin{figure}[H]
\centering
\includegraphics[scale=0.55]{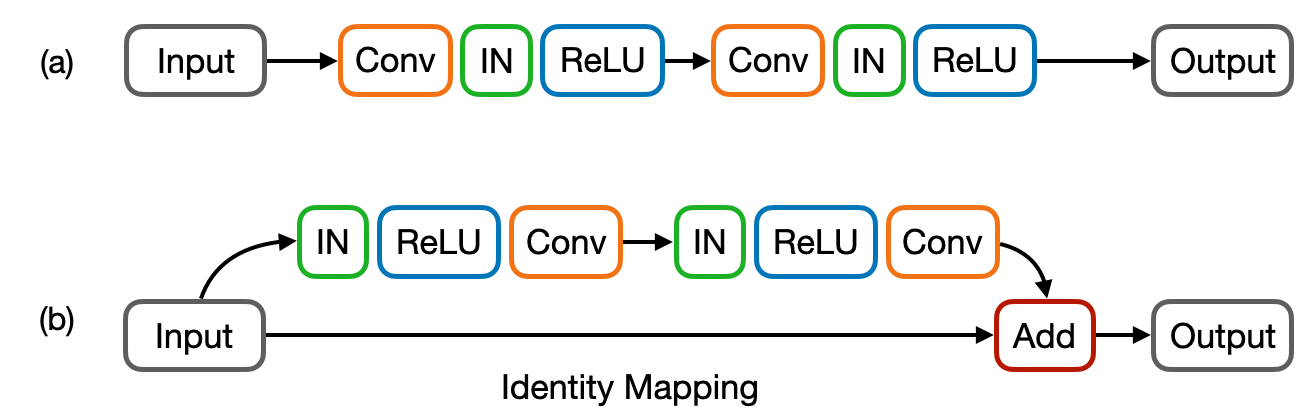}
\caption{Difference between blocks in basic U-Net (a) and Residual U-Net (b). Conv block (orange) corresponds to a convolutional layer with 3x3x3 kernels, IN (green) is instance normalization and ReLU (blue) is a Rectified Linear Unit activation. }
\label{fig:resunet}
\end{figure}

\cite{resnet} is inspired by a ResNet model \cite{resnet} where residual connections were proposed. Adding residual connections is helping with training a deep neural network due to better gradient flow. The only difference between basic U-Net and Residual U-Net is the computation within a convolutional block, which is shown in the Fig. \ref{fig:resunet}.

\vspace{-4mm}
\vspace{-5mm}

Based on our experiments (the detailed results are shown in subsection \ref{sub:exp}), a basic U-Net achieves the best results, and was selected for further exploration. The next optimization was adjusting the encoder depth and optimal selection of the convolution channels. As a baseline, a default U-Net architecture from the nnU-Net framework was used, i.e., the depth of the network was $6$, and the convolution channels at each encoder level were: $32, 64, 128, 256, 320, 320$. Our experiments have demonstrated that increasing the depth of the encoder to 7, and modifying the number of channels to: $64, 96, 128, 192, 256, 384, 512$, further improves the baseline score.

\subsection{Loss function}\label{sub:loss}

Based on the \textit{nnU-Net for Brain Tumor Segmentation}~\cite{introduction_nnunet_brats} paper, the classes present in the label were converted to the three partially overlapping regions: whole tumor (WT) representing classes $1$, $2$, $4$; tumor core (TC) representing classes $1$, $4$; and enhancing tumor (ET) representing the class $4$. The contest leaderboard is computed based on those overlapping regions instead of classes present in the labels. It is beneficial to construct the loss function based on classes used for ranking calculation, thus we designed the output feature map to have three channels (one per class) which at the very end are transformed via the sigmoid activation.

Each region was optimized separately with a sum of binary cross-entropy or Focal loss~\cite{focalloss} (with gamma parameter set to $2$) with the Dice loss~\cite{diceloss}. For Dice loss, its batched variant was used, i.e., Dice loss was computed over all samples in the batch instead of averaging the Dice loss over each sample separately.

\subsubsection{Deep supervision}\label{sub:ds} \cite{ds} is a technique that helps with a better gradient flow by computing loss function on different decoder levels. In this work, we added two additional output heads, marked by green bars on Fig. \ref{fig:unet}. To compute the deep supervision loss, labels were first downsampled using nearest neighbor interpolation to the $(64, 64, 64)$ and $(32, 32, 32)$ spatial shapes such that they match the shapes of additional outputs. For labels $y_i$ and predictions $p_i$ for $i=1,2,3$, where $i=1$ corresponds to the last output head, $i=2$ is the output head on the penultimate decoder level and $i=3$ is before the penultimate, final loss function is computed as follows: \begin{equation} \mathcal{L}(y_1, y_2, y_3, p_1, p_2, p_3) = \mathcal{L}(y_1, p_1) + \frac{1}{2}\mathcal{L}(y_2, p_2) + \frac{1}{4} \mathcal{L}(y_3, p_3).\end{equation}

\subsection{Inference}\label{sub:infer}

During inference, the input volume can have arbitrary size, instead of the fixed patch size $(128, 128, 128)$ as during the training phase. Thus, we used a sliding window inference \footnote{MONAI \href{https://docs.monai.io/en/latest/inferers.html}{sliding window implementation} was used.}, where the window has the same size as the training patch, i.e., $(128, 128, 128)$ and adjacent windows overlap by half the size of a patch. The predictions on the overlapping regions are then averaged with Gaussian importance weighting, such that the weights of the center voxels have higher importance, as in the original nnU-Net paper \cite{introduction_nnunet}.

One of the known ways to improve robustness of predictions is to apply test time augmentations. During inference, we have created eight versions of the input volume, such that each version corresponds to one of eight possible flips along the $x, y, z$ axis combination. Then we run inference for each version of the input volume and transform the predictions back to the original input volume orientation by applying the same flips to predictions as were used for the input volume. Finally, the probabilities from all predictions were averaged.

By optimizing the three overlapping regions (ET, TC, WT) we had to convert them back to the original classes (NCR, ED, ET). The strategy for transforming classes back to the original one is the following: if the WT probability for a given voxel is less than $0.45$ then its class is set to $0$ (background), otherwise if the probability for TC is less than $0.4$ the voxel class is $2$ (ED), and finally if probability for ET is less than $0.45$ voxel has class $1$ (NCR), or \mbox{otherwise $4$ (ET).}

Furthermore, we applied the following post-processing strategy: find ET connected components, for components smaller than 16 voxels with mean probability smaller than $0.9$, replace their class to NCR (such that voxels are still considered part of the tumor core), next if there is overall less than $73$ voxels with ET and their mean probability is smaller than $0.9$ replace all ET voxels to NCR. With such post-processing we avoided the edge case where the model predicted a few voxels with enhancing tumor, but there were not any in the ground truth. Such post-processing was beneficial to the final score as if there were no enhancing tumor voxels in the label, then the Dice score for zero false positive prediction was $1$, and $0$ otherwise.

That methodology was tested on the validations sets from the 5-fold cross-validation. Hyperparameters were selected to yield the highest score combined on all the folds. The threshold value was selected via a grid search method with a step of 0.05 in the range (0.3, 0.7). Similarly, the number of voxels was searched in the range (0, 100) and selected by maximizing score on the 5-fold cross-validation.

\section{Results}\label{ch:results}

\subsection{Implementation}\label{sub:impl}

Our solution is written in PyTorch~\cite{pytorch} and extends NVIDIA's implementation of the nnU-Net. The code is publicly available on the NVIDIA Deep Learning Examples \mbox{GitHub repository} \footnote{\url{https://github.com/NVIDIA/DeepLearningExamples/tree/master/PyTorch/Segmentation/nnUNet}}. Proposed solution is using the NVIDIA NGC PyTorch 21.07 Docker container \footnote{\url{https://ngc.nvidia.com/catalog/containers/nvidia:pytorch}} which allows for the full encapsulation of dependencies, reproducible runs, as well as easy deployment on any system. All training and inference runs were performed with use of \mbox{Mixed Precision~\cite{amp}}, which speeds-up the model and reduces the GPU memory consumption. \mbox{Experiments} were run on NVIDIA DGX A100 \mbox{(8×A100 80 GB) system.}\footnote{\url{https://www.nvidia.com/en-us/data-center/a100}}

\subsection{Training schedule}

Each experiment was trained for $1,000$ epochs using the Adam optimizer~\cite{adam} with three different learning rates: $0.0005$, $0.0007$, $0.0009$ and a weight decay equal to $0.0001$. Additionally, during the first 1000 steps, we used a linear warm-up of the learning rate, starting from 0 and increasing it to the target value, and then it was decreased with a cosine annealing scheduler~\cite{scheduler}. The weights for 3D convolutions were initialized with Kaiming initialization~\cite{init}.

For model evaluation, we used 5-fold cross validation and compared the average of the highest Dice score reached on each of the 5-folds. The evaluation on the validation set was run after every epoch. For each fold, we have stored the two checkpoints with the highest mean Dice score on the validation set reached during the training phase. Then during the inference phase, we ensembled the predictions from stored checkpoints by averaging the probabilities.

\subsection{Experiments}\label{sub:exp}

To select the model architecture, we experimented with three U-Net variants: baseline U-Net~\cite{introduction_unet} which architecture follows the nnU-Net~\cite{introduction_nnunet} architecture heuristic, UNETR~\cite{introduction_unetr} which replaces the U-Net encoder with a Vision Transformer \mbox{(ViT)~\cite{vit}} generalization for the 3D convolutions, and U-Net with autoencoder regularization (SegResNetVAE) which extends U-Net architecture with variational autoencoder (VAE)~\cite{vae} branch for input reconstruction in \mbox{the decoder.}

\begin{table}[H]
\vspace{-3mm}
\centering
\begin{tabular}{|P{2cm}|P{2cm}P{2cm}P{2.5cm}|} \hline
    {Model} & {\text{U-Net}} & {\text{UNETR}} & {\text{SegResNetVAE}}  \\ \hline
    \text{Fold 0} & \textbf{0.9087} & 0.9044 & 0.9086 \\ \hline
    \text{Fold 1} & \textbf{0.9100} & 0.8976 & 0.9090 \\ \hline
    \text{Fold 2} & \textbf{0.9162} & 0.9051 & 0.9140 \\ \hline
    \text{Fold 3} & \textbf{0.9238} & 0.9111 & 0.9219 \\ \hline
    \text{Fold 4} & \textbf{0.9061} & 0.8971 & 0.9053 \\ \hline
    \text{Mean Dice} & \textbf{0.9130} & 0.9031 & 0.9118 \\ \hline
\end{tabular}
\vspace{2mm}
\caption{Averaged Dice scores of ET, TC, WT classes for each 5-folds comparing the baseline U-Net, UNETR, SegResNetVAE models.}
\label{table:res1}
\vspace{-8mm}
\end{table}

Presented results in the Table \ref{table:res1} have shown that baseline U-Net achieves the highest score. Although the score of SegResNetVAE is similar to the plain U-Net, the training time is three times longer compared to U-Net, because of the additional VAE branch. Thus, we decided to select U-Net architecture for further exploration.

In the next phase of experiments we tested various U-Net architecture tweaks: decoder attention~\cite{atten}, deep supervision~\cite{ds}, residual connections~\cite{resnet} and drop block~\cite{drop}. Additionally, we have experimented with the modified loss function with Focal loss~\cite{focalloss} instead of cross-entropy, so that the loss function \mbox{was Focal+Dice.}

The experimental results presented in the Table \ref{table:res2} have shown that the only extension which significantly improves the 5-fold average Dice score over the baseline U-Net (0.9130) was the deep supervision (0.9149).

\begin{table}[H]
\vspace{-3mm}
\centering
\begin{tabular}{|P{2cm}|P{1.6cm}P{1.6cm}P{1.6cm}P{1.6cm}P{1.6cm}P{1.6cm}|} \hline
    {Model} & {\text{baseline}} & {\text{Attention}} & {\text{DS}} & {\text{Residual}} & {\text{DB}} & {\text{Focal}}  \\ \hline
    \text{Fold 0} & 0.9087 & 0.9091 & \textbf{0.9111} & 0.9087 & 0.9096 & 0.9094 \\ \hline
    \text{Fold 1} & 0.9100 & 0.9110 & \textbf{0.9115} & 0.9103 & 0.9114 & 0.9026 \\ \hline
    \text{Fold 2} & 0.9162 & 0.9157 & \textbf{0.9175} & \textbf{0.9175} & 0.9159 & 0.9146 \\ \hline
    \text{Fold 3} & 0.9238 & 0.9232 & \textbf{0.9268} & 0.9233 & 0.9241 & 0.9229 \\ \hline
    \text{Fold 4} & 0.9061 & 0.9061 & \textbf{0.9074} & 0.9070 & 0.9071 & 0.9072 \\ \hline
    \text{Mean Dice} & 0.9130 & 0.9130 & \textbf{0.9149} & 0.9134 & 0.9136 & 0.9133 \\ \hline
\end{tabular}
\vspace{2mm}
\caption{Averaged Dice scores of ET, TC, WT classes for each 5-folds comparing the decoder attention (Attention), deep supervision (DS), residual connections (Residual), drop block (DB) and Focal loss (Focal).}
\label{table:res2}
\vspace{-8mm}
\end{table}

Finally, for the U-Net with deep supervision, we tested the modification of the U-Net encoder. The baseline U-Net architecture follows the architecture heuristic from the nnU-Net~\cite{introduction_nnunet} framework for which the depth of the network was 6, and the convolution channels at each encoder level were: $32, 64, 128, 256, 320, 320$. We experimented with an encoder of depth 7, modified the number of channels to: $64, 96, 128, 192, 256, 384, 512$, and checked the input volume with an additional channel with one-hot encoding for foreground voxels.

\begin{table}[H]
\vspace{-3mm}
\centering
\begin{tabular}{|P{1.7cm}|P{1.7cm}P{1.7cm}P{1.7cm}P{1.7cm}P{1.7cm}|} \hline
    {Model} & {\text{DS}} & {\text{Deeper}} & {\text{Channels}} & {\text{One-hot}} & {\text{D+C+O}}  \\ \hline
    \text{Fold 0} & 0.9111 & \textbf{0.9118} & 0.9107 & 0.9109 & \textbf{0.9118} \\ \hline
    \text{Fold 1} & 0.9115 & 0.9140 & 0.9135 & 0.9132 & \textbf{0.9141} \\ \hline
    \text{Fold 2} & 0.9175 & 0.9170 & 0.9173 & 0.9174 & \textbf{0.9176} \\ \hline
    \text{Fold 3} & \textbf{0.9268} & 0.9256 & 0.9265 & 0.9263 & \textbf{0.9268} \\ \hline
    \text{Fold 4} & 0.9074 & \textbf{0.9079} & 0.9072 & 0.9075 & 0.9076 \\ \hline
    \text{Mean Dice} & 0.9149 & 0.9152 & 0.9150 & 0.9050 & \textbf{0.9156} \\ \hline
\end{tabular}
\vspace{2mm}
\caption{Averaged Dice scores of ET, TC, WT classes for each 5-folds comparing the deep supervision (DS), deeper U-Net encoder, modified number of convolution channels, additional input channel with one-hot encoding for foreground voxels, and all modification applied together (D+C+O) i.e., deeper U-Net with changed number of convolution channels and one-hot encoding channel for foreground voxels.}
\label{table:res3}
\vspace{-8mm}
\end{table}

The results in the Table \ref{table:res3} have shown that applying each of the modifications separately is slightly improving the score over baseline U-Net with deep supervision (0.9149), however if using all modifications together then the score is further improved (0.9156).

Finally, we experimented with a post-processing strategy. It is known from previous BraTS editions that removing small regions with enhanced tumor can be beneficial to the final score. It is so because if there is no enhancing tumor in the label, then the Dice score for zero false positive prediction is $1$, and $0$ otherwise. The best strategy we found for our 5-fold cross-validation is the following: find ET connected components, for components smaller than 16 voxels with mean probability smaller than $0.9$, replace their class to NCR, next if there is overall less than $73$ voxels with ET and their mean probability is smaller than $0.9$ replace all ET voxels to NCR.

\begin{figure}[!t]
\centering
\includegraphics[scale=0.4]{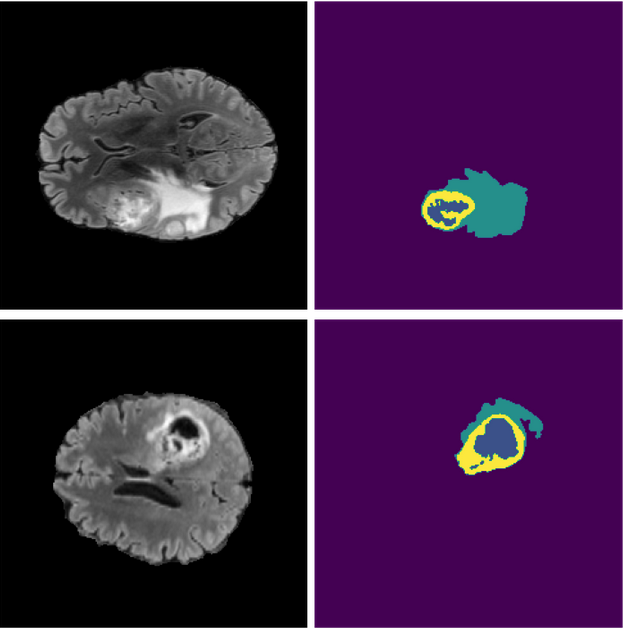}
\caption{Predictions on the challenge validation dataset. On the left column FLAIR modality is visualized while on the right model predictions where the meaning of colors is the following: purple - background, blue - NCR, turquoise - ED, yellow - ET.}
\label{fig:predictions}
\vspace{-4mm}
\end{figure}

\begin{table}[H]
\vspace{-3mm}
\centering
\begin{tabular}{|P{2.5cm}|P{2cm}P{2cm}|} \hline
    {\text{Post-processing}} & {\text{without}} & {\text{with}}  \\ \hline
    \text{Fold 0} & 0.9118 & \textbf{0.9132} \\ \hline
    \text{Fold 1} & 0.9141 & \textbf{0.9142} \\ \hline
    \text{Fold 2} & 0.9176 & \textbf{0.9189} \\ \hline
    \text{Fold 3} & \textbf{0.9268} & \textbf{0.9268} \\ \hline
    \text{Fold 4} & 0.9076 & \textbf{0.9086} \\ \hline
    \text{Mean Dice} & 0.9156 & \textbf{0.9163} \\ \hline
\end{tabular}
\vspace{2mm}
\caption{Averaged Dice scores of ET, TC, WT classes for each 5-folds without and with post-processing.}
\label{table:res4}
\vspace{-8mm}
\end{table}

The best model, i.e., deeper U-Net with deep supervision, modified number of channels and additional input one-hot encoding channel for foreground voxels was the winner of the challenge validation phase. The detailed scores are shown in the \mbox{Table \ref{table:res4}.}

\begin{table}[H]
\centering
\vspace{-1mm}
\begin{tabular}{P{1.2cm}P{1.2cm}P{1.2cm}P{1.2cm}P{1.2cm}P{1.2cm}P{1.2cm}P{1.2cm}P{1.2cm}} \toprule
     {\text{1}} & {\text{2}} & {\text{3}} & {\text{4}} & {\text{5}} & {\text{6}} & {\text{7}} & {\text{8}} & {\text{9}} \\ \midrule
    \textbf{0.267} & 0.272 & 0.287 & 0.289 & 0.298 & 0.305 & 0.306 & 0.312 & 0.316 \\ \bottomrule
\end{tabular}
\vspace{2mm}
\caption{Top 9 normalized statistical ranking scores for BraTS21 validation phase.}
\label{table:res4}
\end{table}

\section{Conclusions}\label{ch:discussion}

We have experimented with various U-Net variants (basic U-Net \cite{introduction_unet}, \mbox{UNETR \cite{introduction_unetr}}, SegResNetVAE \cite{introduction_myronenko}, Residual U-Net \cite{resnet}, and Attention U-Net \cite{atten}), architecture modifications and training schedule tweaks like: deep supervision \cite{ds}, drop block \cite{drop}, and Focal loss \cite{focalloss}. Based on our experiments, U-Net with deep supervision yields the best results which can be further improved by adding an additional input channel with one-hot encoding for foreground, increasing encoder depth together with a number of convolutional channels and designing a post-processing strategy.

\nocite{*}
\bibliographystyle{splncs04}
\bibliography{bib.bib}

\end{document}